\documentclass[apjl]{emulateapj}
\usepackage{apjfonts}

\shorttitle{MISSING BARYONS IN CLUSTERS AND GROUPS}
\shortauthors{CAVALIERE \& LAPI}
\journalinfo{Accepted by The Astrophysical Journal.}

\begin{document}

\title{Missing Baryons, from Clusters to Groups of Galaxies}
\author{A. Cavaliere\altaffilmark{1} and A. Lapi\altaffilmark{1,2}}
\altaffiltext{1}{Astrofisica, Dip. Fisica, Univ. `Tor Vergata',
Via Ricerca Scientifica 1, 00133 Roma, Italy.}
\altaffiltext{2}{Astrophysics Sector, SISSA/ISAS, Via Beirut
2-4, 34014 Trieste, Italy.}

\begin{abstract}
From clusters to groups of galaxies, the powerful
bremsstrahlung radiation $L_X$ emitted in X rays by the
intracluster plasma is observed to decline sharply with
lowering virial temperatures $T$ (i.e., at shallower depths of
the gravitational wells) after a steep local $L_X - T$
correlation; this implies increasing scarcity of diffuse
baryons relative to dark matter, well under the cosmic
fraction. We show how the widely debated issue concerning these
`missing baryons' is solved in terms of the thermal and/or
dynamical effects of the kinetic (at low redshifts $z$) and
radiative (at high $z$) energy inputs from central active
galactic nuclei, of which independent evidence is being
observed. From these inputs we compute shape and $z$-evolution
expected for $L_X- T$ correlation which agree with the existing
data, and provide a predictive pattern for future observations.
\end{abstract}
\keywords{galaxies: active nuclei -- galaxies: clusters --
radiosources -- X rays: galaxies: clusters}

\section{Introduction}

Groups and clusters of galaxies with their masses in the range
$M\sim 10^{13} - 10^{15}\, M_{\odot}$ constitute the largest
virialized structures in the Universe. They are dominated by
the dark matter (DM) component, and are built up in a
hierarchical sequence by gravitational infall and merging of
smaller structures. Thus gravitational potential wells are set
up with vast virial radii $R\sim 0.2-3$ Mpc, and large depths
gauged by member velocity dispersions $\sigma^2 \sim G\, M/5\,
R \sim (0.2 - 2\times 10^3 ~\mathrm{km~s}^{-1})^2$.

Such wells might be expected to also drag in and contain inside
$R$ all the baryons originally associated with the DM in making
up the cosmic density ratio $\rho_b/\rho \approx 0.17$. Apart
from a minor fraction locked up into galactic stars, these
ought to be just reshuffled to constitute with the neutralizing
electrons a diffuse intracluster plasma (ICP) in thermal and
gravitational equilibrium at temperatures close to the virial
values $kT \approx m_p\, \sigma^2/2 \sim 1 - 10$ keV.

On the other hand, the actual number density $n = \rho_b/m_p$
of the baryons (mostly protons with mass $m_p$) in the ICP is
probed from its strong X-ray emissions $L_X \propto n^2\, R^3\,
T^{1/2} \sim 10^{42} - 10^{46}$ erg s$^{-1}$ produced by
electron bremsstrahlung radiation. It is found that the
baryonic fraction $f_b$ is particularly low in poor clusters
and groups despite these wells being, if anything, deeper (more
`concentrated', see Navarro et al. 1997) than gauged from
$\sigma^2$ or $2\,kT$.

The finding stems from the correlation observed locally between
$L_X$ and $T$. Its generic form is outlined by the relation
\begin{equation}
L_X\propto f^2_b \; \rho^{1/2}\; T^2~,
\end{equation}
provided by the brems emission law coupled with the virial
scaling $kT \propto \rho \, R^2$ enforced by the DM gravity
(Kaiser 1986). But the actual correlation differs sharply from
the  shape $L_X\propto T^2$ expected for wells filled up to a
constant $f_b$; if anything, this ought to be bent upwards  for
groups with $kT < 2$ keV where line emission adds to the brems
continuum. Instead, poor clusters and groups radiate much less
than so expected, with the actual correlation bent
\emph{downward} to $L_X\propto T^3$ and steeper (see Osmond \&
Ponman 2004).

These low radiation levels witnessing to scarcity of
electrons-baryons have stirred a wide debate (see Evrard \&
Henry 1991; Ponman et al. 1999; Cavaliere et al. 2002; Lapi et
al. 2005; Borgani et al. 2006; Bregman et al. 2007) over the
mystery of the baryons missing from the ICP: are they lost
somewhere within the wells, or outflown, or just limitedly
infallen? Here we show that the second alternative is the
fitting one, with some help from the third.

\begin{figure*}
\epsscale{1.}\plotone{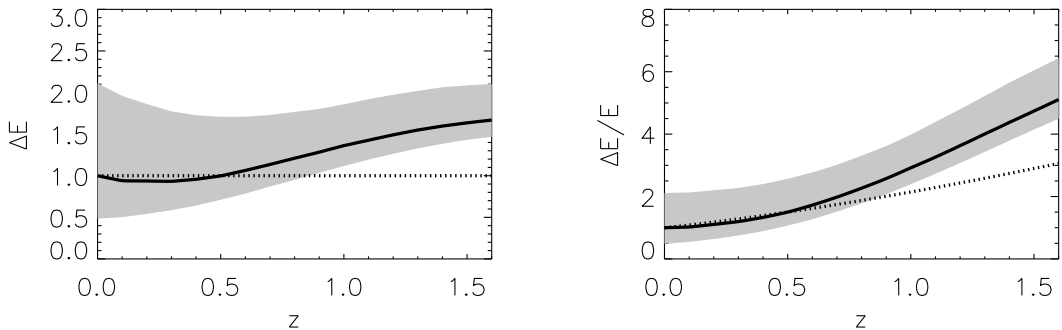} \caption{Left: the
\emph{solid} line shows the redshift evolution of the average
energy input $\Delta E$ (normalized to the present) from AGNs
(see Eq.~[2]), with the \emph{shaded} area illustrating the
uncertainties (see text for details); the \emph{dotted} line
marks for reference the non-evolving case. Right: same for the
ratio $\Delta E/E$ of the energy input to the ICP binding
energy.}
\end{figure*}

\section{Cooling vs. heating}

In sorting out these alternatives it helps to note that low
emissivity $\mathcal{L}_X \propto n^2\, T^{1/2}$ from the
current ICP goes along with enhanced adiabat $K \propto kT\,
n^{-2/3}\propto T^{5/3}\, \mathcal{L}_X^{-1/3}$ or excess
specific entropy proportional to $ln{K}$, as in fact is
measured (see Ponman et al. 2003; Piffaretti et al. 2005; Pratt
et al. 2006).

Baryon sinks  may occur within the wells due to radiative
cooling, that affects mostly the densest and lowest entropy
fractions of the ICP. Thereby these tend to lose their pressure
support and further condense, radiate and cool even more,  and
so engage in the classic catastrophic course ending up into
formation of plenty new stars (White \& Rees 1978; Blanchard et
al. 1992). What is left  after a Hubble time is the ICP
fraction originally in a hot dilute state, exhibiting a high
residual entropy today (Voit 2005).

This indirect  mechanism faces limitations, however. For
example, cooling would steepen the local $L_X - T$ correlation
to a shape $L_X \propto T^3$ only if it proceeded unscathed for
a Hubble time throughout the hectic sequence of hierarchical
mergers that build up a cluster of today (Voit 2005; for
related evidence see Vikhlinin et al. 2007). Even so, the
corresponding height still would fall short of the observed
luminosity or entropy levels, unless so many stars condensed as
to exceed the observational limits (Muanwong et al. 2002).

Finally, cooling does not even  dominate  the very `cooling
cores' at the center of the ICP; these are observed in many
clusters to feature enhanced emissions and short cooling times,
but their temperatures are bounded by $T_c \ga T/3$ (Molendi \&
Pizzolato 2001; Peterson \& Fabian 2006) at variance with the
catastrophic trend intrinsic to cooling. So the latter must be
offset by inputs of energy into the ICP, most likely from
moderately active nuclei (AGNs) of central member galaxies
(Binney \& Tabor 1995; Voit \& Donahue 2005).

But then such internal inputs can \emph{directly} produce
heating and outflows of the ICP so as to substantially lower
$L_X$ or equivalently raise the entropy. To affect the ICP at
large, inputs of a few  keVs per particle are required, and are
easily provided by powerful AGNs energized by accreting
supermassive black holes (BHs) in member galaxies (Wu et al.
2000; Cavaliere et al. 2002; Nath \& Roychowdhury 2002; Lapi et
al. 2005). These not only effectively control the central cool
cores within $100$ kpc, but also can raise $K$ by some $10^2$
keV cm$^2$ over much larger ICP masses (Lapi et al. 2005). In
fact, imprints of such inputs in action are directly observed
in the ICP out to $r\approx 3\times 10^2$ kpc, where they
excavate extensive cavities or launch far-reaching blastwaves
(B\^{i}rzan al. 2004; Nulsen et al. 2005; Forman et al. 2005;
Cavaliere \& Lapi 2006).

Taking up from Lapi et al. (2005), we show next how these AGN
inputs provide a unified key to both the steep \emph{shape} of
the local $L_X - T$ relation and the apparently non-monotonic
evolution of its \emph{height}.

\section {AGN inputs, kinetic vs. radiative}

What kind of AGNs are most effective in this context, depends
on $z$ and on the BH accretion rates $\dot m$ (Eddington
units).

At $\dot{m}\la 5\%$ and low $z\la 0.5$ the dominant component
to the outputs is constituted by `kinetic power', in the form
of high speed conical winds and narrow relativistic jets
associated with radio emissions (see Churazov et al. 2005;
Blundell \& Kuncic 2007; Merloni \& Heinz 2007; Heinz et al.
2007). Such outputs already in kinetic form end up with
coupling a substantial energy fraction $f_k\approx 1$ to the
surrounding medium, interstellar or ICP.

At higher $\dot{m}$ the radiative activity is held to be
dominant (see Churazov et al. 2005), and is also well known to
evolve strongly with $z$. So this is expected to take over in
affecting the ICP despite the weaker photon-particle energetic
coupling, bounded to levels $f_r \approx v/2c\approx 0.05$ by
momentum conservation setting an outflow speed $v$. We stress
that values of that order effectively account in terms of AGN
feedback not only for the local shape of the $L_X-T$
correlation in clusters and groups that started up the present
investigation, but also for several other observables: the
galaxy stellar mass and luminosity functions (Springel et al.
2005); the luminosity functions of quasars and the mass
distribution of relic BHs (Hopkins et al. 2006; Lapi et al.
2006); the correlation between these masses and the velocity
dispersions in the host bulges (Vittorini et al. 2005).

\begin{deluxetable}{ccccc}
\tablecaption{Estimated contributions to feedback}
\tablehead{\colhead{} & \multicolumn{3}{c}{$R_{k,r}$} & \colhead{}\\
\cline{2-4}\\
\colhead{Comp.} & \colhead{r. quiet ($0.9$)} & \colhead{r. loud
($0.1$)} & \colhead{Total} & \colhead{$f_{k,r}$}} \startdata
rad. & $1$  & $1/2$ & $0.95$ & $\la 0.05$\\
kin. & $0$  & $1/2$ & $0.05$ & $1$
\enddata
\end{deluxetable}

On the other hand, velocities up to $v\sim 0.1 - 0.2\, c$ and
metals spread out to some $10^2$ kpc are widely observed around
radioquiet quasars, indicating the action of efficient
radiation-driven outflows or blastwaves (e.g., Pounds \& Page
2006, Stockton et al. 2006). The radiation thrust may involve
absorbtion in many atomic lines, or Thomson scattering of the
continuum in the galactic plasma. Based on the latter, King
(2003) has computed in detail outflows starting from the
Compton-thick vicinities of the central BH and continuously
accelerated to high speeds, that drive powerful blastwaves
propagating into the outskirts of the host galaxy and beyond.
At that point the Thomson optical depth retains values of
several $10^{-2}$, having decreased as $r^{-0.5}$ or slower in
the hot galactic plasma with its flatter distribution relative
to DM's.

\begin{figure*}
\epsscale{.7}\plotone{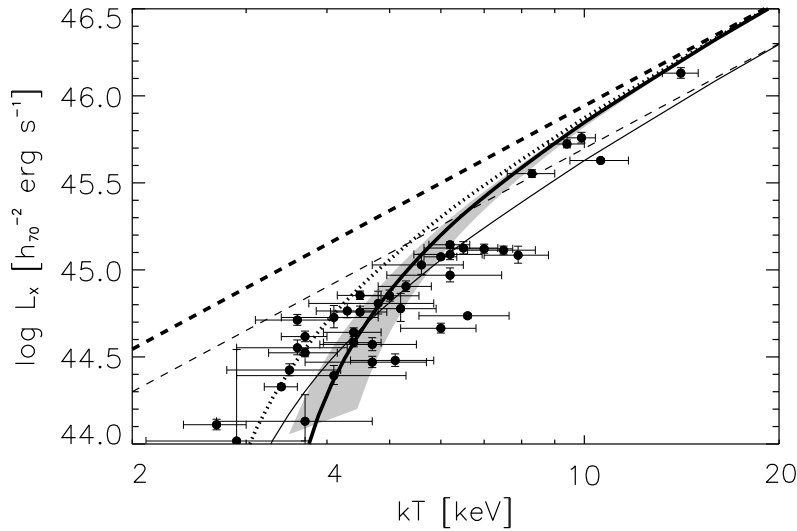} \caption{The $L_X-T$
relation at different redshifts $z$. The data (\textit{filled
circles}) are from the high-$z$ (average $z\approx 0.7$) sample
by Branchesi et al. (2007). The \emph{thick} lines refer to $z
= 0.7$: the \emph{dashed} one is the relation in the absence of
energy inputs; the \emph{solid} one shows the model result we
find from Eq.~(4) on using the $z$-dependence of $\Delta E/E$
after Eqs.~(2) and (3); the \emph{dotted} one is for a
non-evolving $\Delta E$. For comparison, the \emph{thin} lines
refer to $z=0$: the \emph{dashed} one is the relation in the
absence of energy inputs, and the \emph{solid} one shows the
model result for $z=0$, that provides a good fit to the local
data (Lapi et al. 2005, see their Figure 3).}
\end{figure*}

The integrated input levels in these two regimes are related to
a first approximation by the simple `golden rule' $f\, \dot{m}
\approx $ const, cf. Churazov et al. (2005). But their
$z$-dependencies differ strongly as stressed by Merloni \&
Heinz (2007) and Heinz et al. (2007), with the kinetic power
roughly constant (or slowly decreasing) out to $z \approx 0.5$,
and the radiative power strongly increasing out to $z \approx
2.5$ if limited in its effects by weaker coupling.

These two components involve an activity fraction $R_{k,r}$ for
radio loud or radio quiet AGNs, and contribute energies
$W_{k,r}$ which combine after the reckoning given in Table 1 to
yield the overall input
\begin{equation}
\Delta E(z) = R_k \, f_k\, W_k(z) + R_r \, f_r \, W_r(z)~.
\end{equation}
As to $W_{k,r}$ we adopt the evaluations provided by Merloni \&
Heinz (2007); note that the kinetic contribution is mainly
inferred from radio emission and from lack of X rays observed
around radio loud AGNs, while the radiative contribution is
measured mainly through the emissions from IR to X rays of
radio quiet ones. The evolution of the overall $\Delta E(z)$ is
reported in Fig.~1 (left) with uncertainties (shaded area)
dominated by the jet beaming factors entering the determination
of $W_k$.

Consider now that any action of the inputs $\Delta E$ on the
ICP is to be gauged against its unperturbed total energy $E$
(thermal plus gravitational), with modulus scaling as $E\propto
kT\, M_b\propto f_b\,\rho^{-1/2}\, T ^{5/2}$; the dependencies
on $z$ and $T$ are given in full by
\begin{equation}
E(z,T)\approx 1.5\times 10^{63}\, {f_b\over 0.12}\,
{H(0)\, \Delta_{\rm v}^{1/2}(0)\over H(z)\, \Delta_{\rm v}^{1/2}(z)}\,
\left({kT\over 6~ \mathrm{keV}}\right) ^{5/2}~\mathrm{erg}~.
\end{equation}
Here the DM density $\rho \propto H^2 \, \Delta_v$ internal to
a virializing cluster or group has been taken from the standard
`top hat' collapse model for structure formation, in terms of
the running Hubble parameter $H(z)$ and the contrast
$\Delta_{\rm v}(z)$ at virialization (see Peebles 1993); in the
Concordance Cosmology the resulting behavior for $z\la 1$ is
close to $\rho(z)\propto (1+z)$.

The above expression for $E$ strictly holds for an isothermal
ICP filling an isothermal DM potential well; however, for a
well shape after Navarro et al. (1997) and/or a polytropic
distribution of the ICP with the appropriate index $\Gamma
\approx 1.1- 1.2$ the result is not significantly altered; in
particular, it is unaltered the slow decrease for $z \ga 0.5$
due to smaller masses associated with earlier potential wells.
Normalization is made to the value $f_b = 0.12$ provided by
X-ray and Sunyaev-Zel'dovich observations of ICP in rich
clusters (see LaRoque et al. 2006; Ettori et al. 2006); the
discrepancy from the value $0.14$ (expected on subtracting the
stellar component from the cosmic value $0.17$) is likely due
to the parallel, second order effect of the AGN feedback
causing external preheating and entropy rise of the gas, which
limit its infall into the gravitational wells as discussed by
Lapi et al. (2005). These authors stress that preheating by
itself would also cause wider distributions of the ICP in poor
clusters and groups; but in these smaller potential wells the
effect is actually offset by their higher concentration,
resulting in nearly invariant profiles as observed, e.g., by
Pratt \& Arnaud (2003) and by Pratt et al. (2006).

Thus the effective input at each $z$ is given by the ratio
$\Delta E/E$, that we represent in Fig.~1 (right) after
normalizing it to the value needed to fit the local $L_X-T$
relation; this corresponds to coupling levels $f_r\approx
0.05$, discussed above in this Section.

\section{Results}

Given this input, we compute the resulting $L_X$ on extending
to higher $z$ the approach that yields a fitting shape for the
local $L_X -T$ all the way from $10$ to $1/2$ keV (see Lapi et
al. 2005, their Fig.~3). So we substantiate Eq.~(1) with the
baryonic fraction $f_b\, [1- \Delta E/2E]$ affected by internal
AGNs, to read
\begin{equation}
L_X(z,T)\propto f_b\, H(z)\; \Delta_{\rm v}^{1/2}(z)\;
\left[1- {\Delta E(z)\over 2\, E(z,T)}\right]^2\, T^2 ~.
\end{equation}

\begin{figure*}
\epsscale{.7}\plotone{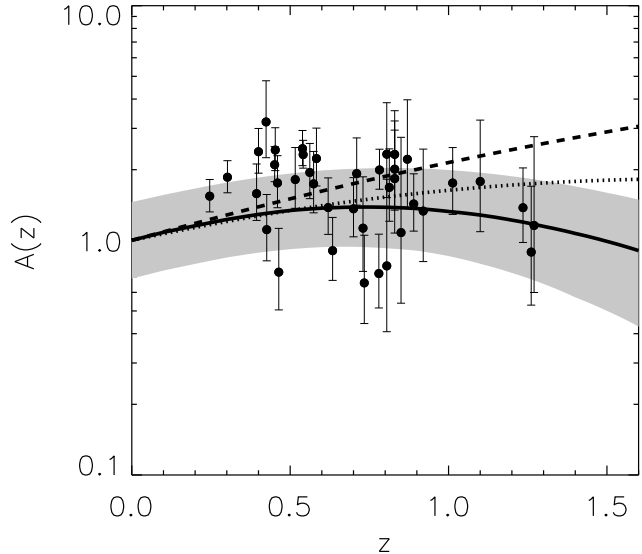}\caption{Redshift evolution
of $\mathcal{A}(z)=L_X(z)/L_X(0)$, see Section 4. The
\emph{solid} line is the evolution  from using Eq.~(4) with
$\Delta E/E$ as in Fig.~1, right panel; the \emph{dotted} line
is for a constant input $\Delta E$; the \emph{dashed} line
represents the scaling expected in the  absence of energy
inputs. Data (\textit{filled circles}) are from  Branchesi et
al. (2007).}
\end{figure*}

The factor in square brackets is the simple, converging result
obtained from detailed numerical modeling; specifically, we use
two oppositely extreme models of thermal outflows and of
dynamical ejection of the ICP out of the DM potential wells,
caused by finite perturbances (blastwaves) driven by AGNs, see
Lapi et al. (2005). The factor $1/2$ reflects the close
equipartition of kinetic and thermal energy in blasts with Mach
numbers $\mathcal{M}\approx 1.5 - 2$ such as involved in this
context. The prefactor $H(z)\, \Delta_{\rm v}^{1/2}(z)$ again
expresses $\rho^{1/2}(z)$.

Fig.~2 shows two snapshots of the $L_X-T$ relation predicted
after such modeling. The one at $z=0$ (thin solid line)
steepens toward low temperatures due to the $T$ dependence of
$E$ given by Eq.~(3); it provides a good fit to the local data,
as discussed by Lapi et al. (2005). For $z\ga 0.5$ we expect
the $L_X-T$ relation (see the thick solid line) to
\emph{steepen} yet relative to that at $z = 0$, due to the
increase of $\Delta E/E$ with $z$; such an evolved relation
agrees with the current observational evaluations, including
the precise ones at high $T$ provided by Branchesi et al.
(2007). Note that a constant $\Delta E$ would yield the dotted
line that systematically overarches most data points.

Fig.~3 shows our prediction for the evolution in rich clusters
of the X-ray luminosity at a given $T$, expressed as
$\mathcal{A}(z)\equiv L_X(z)/L_X(0)$; the data by Branchesi et
al. (2007) have been reported with the same temperatures and
local $L_X-T$ relation they adopt.

Our result agrees with the existing data, with their
non-monotonic \emph{trend} currently just looming out and their
large \emph{scatter}. In our modeling scatter is mainly
contributed by variance in the evaluations of the jet beaming
factors (see Merloni \& Heinz 2007), conceivably reflecting
physical variations. As to trend, our run of $\mathcal{A}(z)$
predicts a clear non-monotonic pattern, with a rise out to $z
\approx 0.5$ and then a decrease to higher $z$. The key feature
to the slow \emph{rise} is that the kinetic power $W_k (z)$,
with its lack of strong evolution, can barely chase the
cosmological increase of all internal densities out to $z
\approx 0.5$. But at higher $z$ the evolution is curbed or
\emph{reversed} by the radiative input emerging; this occurs by
virtue of its strongly positive evolution common to all
radiative AGN activities from IR to X rays, despite the weaker
coupling but with some help from the decrease of $E(z)$.

\section{Discussion and conclusions}

The results of our computations yield a slow, non-monotonic
pattern of $L_X(z)$ at given $T$. This stems from  basic
features of the internal AGN outputs, that comprise two
different components: kinetic and radiative, combining as
follows.

i) Kinetic power at low $\dot{m}$ and $z$. An extreme
interpretation attributes this to the Blandford \& Znajek
(1977) mechanism for extraction of BH rotational energy from
the large reservoir accrued by past accretion events. Low rates
$\dot{m}$ (likely from trickling accretion related to cooling
in the host galaxy) just provide enough material  to hold in
the accretion disk the magnetic field that threads the BH
horizon and induces significant outward Poynting flux.

ii) Radiative power, prevailing at higher $z$ and $\dot{m}$.
This is widely held (Springel et al. 2005; Cavaliere \& Menci
2007; also Conselice 2007) to be driven by violent galaxy
interactions and mergers; their rates increase sharply for
$z\ga 0.8$ in the Concordance Cosmology, so as to drive large
accretion rates onto the central BHs.

Besides interpretations, the fact stands that the two processes
with their different couplings nearly match at $z\approx 0.5$,
in agreement with the golden rule. At lower $z$ the kinetic
mode is granted a leading edge by its stronger coupling with
the ICP; for $z\ga 0.5$, instead, the radiative mode with its
weaker coupling takes over by virtue of its strong evolution.

To conclude, we submit the missing baryons from the ICP of poor
clusters and groups to be explained in terms of the energy
feedback from internal AGNs, a straight extension of the inputs
that limit the cooling cores. In closer look, we predict for
$L_X - T$ a non-monotonic pattern primarily reflecting the two
different types of evolution in the AGN activities: closely
constant as for the kinetic component, and strongly positive
for the radiative one. At given $T$, the former just slows down
the rise of $L_X(z)$ driven by the cosmogonic density increase,
whilst the latter with $f_r\approx 0.05$ can reverse the trend
sharply. While for the body of the current data with their
scatter the  gross average may still be formally compatible
with a monotonic rise (Pacaud et al. 2007), a non-monotonic
pattern is already looming out from luminous high-$z$ clusters,
consistent with our prediction.

In a wider perspective, we argue that the X-ray emission
$L_X(z)$ can independently probe the evolution of the kinetic
power activity mainly related to radio loud AGNs. We submit
that the independent data concerning $L_X$ already indicate for
this component a nearly constant, if not a weakly
\textit{negative} evolution for low $z\la 0.5$; this
constitutes an emerging feature of the kinetic power, with a
precedent only in the lack of evolution (see Caccianiga et al.
2002) of the BL Lac Objects, themselves kinetically loud
sources. The link we establish between $L_X(z) $ and $W_k(z)$
will provide complementary information to direct statistics of
weak radio sources (see discussion by De Zotti et al. 2005),
which is hindered by incompleteness and by confusion from
diffuse galactic contributions.


\begin{references}

\reference{}Binney, J., \& Tabor, G. 1995, MNRAS, 276, 663

\reference{}B\^{i}rzan, L., Rafferty, D.A., McNamara, B.R.,
Wise, M.W., \& Nulsen, P.E.J. 2004, ApJ, 607, 800

\reference{}Blanchard, A., Valls-Gabaud, D., \& Mamon, G.,
1992, A\&A, 264, 365

\reference{}Blandford, R.D., \& Znajek, R. L., 1977, MNRAS,
179, 433

\reference{}Blundell, K.M., \& Kuncic, Z. 2007, 668, L103

\reference{}Borgani, S., et al. 2006, MNRAS, 367, 1641

\reference{}Branchesi, M., Gioia, I.M., Fanti, C., \& Fanti, R.
2007, A\&A, 472, 739

\reference{}Bregman, J.N. 2007, ARA\&A, 45, 221

\reference{}Bryan, G.L., \& Voit, G.M. 2005, Royal Soc. of
London Transactions Ser. A, 363, 715

\reference{}Caccianiga, A., et al. 2002, ApJ, 566, 181

\reference{}Cavaliere, A., \& Menci, N. 2007, ApJ, 664, 47

\reference{}Cavaliere, A., \& Lapi, A. 2006, ApJ, 647, L5

\reference{}Cavaliere, A., Lapi, A., \& Menci, N. 2002, ApJ,
581, L1

\reference{}Churazov, E. et al., 2005, MNRAS, 363, L91

\reference{}Conselice, C.J. 2007, Scientific American, 296, 34

\reference{}De Zotti, G., et al. 2005, A\&A, 431, 893

\reference{}Di Matteo, T., Springel, V., \& Hernquist, L. 2005,
Nature, 433, 604

\reference{}Evrard, A.E., \& Henry, J.P. 1991, ApJ, 383, 95

\reference{}Forman, W., et al. 2005, ApJ, 635, 894

\reference{}Heinz, S., Merloni, A., \& Schwab, J. 2007, ApJ,
658, L9

\reference{}Hopkins, P.F., et al. 2006, ApJ, 652, 864

\reference{}Kaiser, N. 1986, MNRAS, 222, 323

\reference{}King, A.R. 2003, ApJ, 596, L27

\reference{}Lapi, A., et al. 2006, ApJ, 650, 42

\reference{}Lapi, A., Cavaliere, A., \& Menci, N. 2005, ApJ,
619, 60

\reference{}LaRoque, S., et al. 2006, ApJ, 652, 917

\reference{}Merloni, A., \& Heinz, S. 2007, in Black Holes from
Stars to Galaxies - Across the Range of Masses, ed. V. Karas
and G. Matt (Cambridge, UK: Cambridge Univ. Press), 65

\reference{}Molendi, S., \& Pizzolato, F. 2001, ApJ, 560, 194

\reference{}Muanwong, O., Thomas, P., Kay, S.T., \& Pearce,
F.R. 2002, MNRAS, 336, 527

\reference{}Nath, B.B., \& Roychowdhury, S. 2002, MNRAS, 333,
145

\reference{}Navarro, J.F., Frenk, C.S., \& White, S.D.M. 1997,
ApJ, 490,493

\reference{}Nulsen, P.E.J., McNamara, B.R., Wise, M.W., \&
David, L.P. 2005, ApJ, 628, 629

\reference{}Osmond, J.P.F., \& Ponman, T.J. 2004, MNRAS, 350,
1511

\reference{}Pacaud, F., et al. 2007, MNRAS, 382, 1289

\reference{}Peebles, P.J.E. 1993, Principles of physical
cosmology, Princeton: Princeton Univ. Press

\reference{}Peterson, J.R., \& Fabian, A.C. 2006, PhR, 427, 1

\reference{}Piffaretti, R., Jetzer, Ph., Kaastra, J.S., Tamura,
T. 2005, A\&A, 433, 101

\reference{}Ponman, T.J., Sanderson, A.J.R., \& Finoguenov, A.
2003, MNRAS, 343, 331

\reference{}Ponman, T.J., Cannon, D.B., \& Navarro J.F. 1999,
Nature, 397, 135

\reference{}Pounds, K.A., \& Page, K.L. 2006, MNRAS, 372, 1275

\reference{}Pratt, G. W., Arnaud, M., \& Pointecouteau, E.
2006, A\&A, 446, 429

\reference{}Pratt, G. W., \& Arnaud, M. 2003, A\&A, 408, 1

\reference{}Springel, V., et al. 2005, Nature, 435, 629

\reference{}Stockton, A., et al. 2006, ApJ, 638, 635

\reference{}Vikhlinin, A., et al. 2007, in Heating vs. Cooling
in Galaxies and Clusters of Galaxies, ed. H. B\"{o}hringer,
G.W. Pratt, A. Finoguenov, and P. Schuecker (Berlin: Springer),
48

\reference{}Vittorini, V., Shankar, F., \& Cavaliere, A. 2005,
MNRAS, 363, 1376

\reference{}Voit, G.M. 2005, AdSpR, 36, 701

\reference{}Voit, G.M., \& Donahue, M. 2005, ApJ, 634, 955

\reference{}White, S.D.M., \& Rees, M.J. 1978, MNRAS, 183, 341

\reference{}Wu, K.K.S., Fabian, A.C., \& Nulsen, P.E.J. 2000,
MNRAS, 318, 889

\end{references}
\end{document}